\documentclass[preprint,aps,a4paper]{revtex4-1}
\usepackage{mathrsfs}
\usepackage{graphicx}
\usepackage{multirow}
\usepackage{makecell}
\usepackage{booktabs}
\usepackage{lineno}
\usepackage{xcolor}
\usepackage[normalem]{ulem}
\begin{document}

     \title{Coexistence of High Temperature Superconductivity and Antiferromagnetic Order in a Cuprate with Multiple Hole Fermi Pockets}

     \author{Xiangyu Luo$^{1,\dag}$, Yinghao Li$^{1,2,\dag}$, Hao Chen$^{1,2,\dag}$, Yiwen Chen$^{1,2}$, Jumin Shi$^{1,2}$, Taimin Miao$^{1,2}$, Bo Liang$^{1,2}$, Wenpei Zhu$^{1,2}$, Neng Cai$^{1,2}$, Xiaolin Ren$^{1,2}$, Yingjie Shu$^{1,2}$, Chaohui Yin$^{1,2}$, Jiuxiang Zhang$^{1,2}$, Chengtian Lin$^{3}$, Shenjin Zhang$^{4}$, Zhimin Wang$^{4}$, Fengfeng Zhang$^{4}$, Feng Yang$^{4}$, Qinjun Peng$^{4}$, Zuyan Xu$^{4}$, Guodong Liu$^{1,2,5}$, Xintong Li$^{1,2,5}$, Hanqing Mao$^{1,2,5}$, Tao Xiang$^{1,2,5,6}$, Lin Zhao$^{1,2,5,*}$ and X. J. Zhou$^{1,2,5,*}$}

\affiliation{
   \\$^{1}$Beijing National Laboratory for Condensed Matter Physics, Institute of Physics, Chinese Academy of Sciences, Beijing 100190, China.
   \\$^{2}$University of Chinese Academy of Sciences, Beijing 100049, China.
   \\$^{3}$Max Planck Institute for Solid State Research, Heisenbergstrasse 1, D-70569 Stuttgart, Germany.
   \\$^{4}$Technical Institute of Physics and Chemistry, Chinese Academy of Sciences, Beijing 100190, China.
   \\$^{5}$Songshan Lake Materials Laboratory, Dongguan, Guangdong 523808, China.
   \\$^{6}$Beijing Academy of Quantum Information Sciences, Beijing 100193, China.
   \\$^{\dag}$These authors contributed equally to this work.
   \\$^{*}$Corresponding author: LZhao@iphy.ac.cn, XJZhou@iphy.ac.cn
}

	\date{\today}
	
	\maketitle
	
	\newpage

    {\bf  The intricate relationship between high temperature superconductivity and antiferromagnetic order in cuprates, and the fundamental origin of electron pairing remain open questions. By utilizing high-resolution laser-based spatially-resolved angle-resolved photoemission spectroscopy, we investigate the seven-layer $\mathbf{Bi_{2}Sr_{2}Ca_{6}Cu_{7}O_{18+\delta}}$ (Bi2267) and identify a cuprate system that consists of multiple hole Fermi pockets. The observed Fermi pockets exhibit pronounced momentum-, temperature- and Fermi surface-dependent energy gaps. Crucially, high temperature superconductivity with a critical temperature ($T_{\mathrm{c}}$) of $\sim75$\,K emerges in a system with multiple Fermi pockets and the presence of strong antiferromagnetic order and correlations. In particular, substantial electron pairing is observed along the Fermi pocket with an energy gap up to $\sim$42\,meV in lightly-doped CuO$_{2}$ planes ($p\sim0.05$). These findings challenge the conventional understanding of the roles of the nodal and antinodal electronic states in driving high-temperature superconductivity. They show that superconductivity and antiferromagnetism can coexist in a cuprate with multiple Fermi pockets, offering further insights into the pairing mechanism in cuprate superconductors.}

    \vspace{3mm}

\noindent{\bf Introduction}

    Despite nearly four decades of intensive research, the fundamental mechanism underlying high temperature superconductivity in cuprates remains a major challenge in condensed matter physics\cite{PALee2006,BKeimer2015}. A key debate centers on the nature of the pairing glue, the microscopic interaction responsible for the high superconducting critical temperature ($T_{\text{c}}$). In undoped CuO$_{2}$ planes, the hybridization between Cu 3d$_{x^{2}-y^{2}}$ and O 2p orbitals generate strong in-plane superexchange interactions, stabilizing long-range antiferromagnetic (AFM) order\cite{DVaknin1987,JMTranquada1988,EManousakis1991,HEskes1993}. In the usual electronic phase diagram, such a long-range AFM order is rapidly suppressed with slight hole doping, followed by the emergence of superconductivity at higher doping levels\cite{BKeimer2015,CWChu2015}. Within the superconducting doping regime, antiferromagnetic spin fluctuations become pronounced in the underdoped region and persist even into the heavily overdoped region\cite{PCDai1999,RColdea2001,HAMook2002,SMHayden2004,SWakimoto2004,BVignolle2007,JMTranquada2007}. Theoretical studies have considered the spin fluctuations as a prime candidate of pairing glue in generating high temperature superconductivity\cite{DJScalapino1986,PMonthoux1991,DJScalapino1995,TMoriya2000,ArAbanov2008,DJScalapino2012}. These underscore the intricate interplay between antiferromagnetism and superconductivity in cuprate superconductors.

    The extensive studies have also established a general electronic picture of doping the parent Mott insulators of cuprates\cite{ADamascelli2003,TYoshida2003,KMShen2004,KMShen2005,PALee2006,TYoshida2006,MHashimoto2008,YYPeng2013,QGao2020}. Upon slight hole doping of the parent compound, the highest occupied electronic states first emerge near the $(\pi/2,\pi/2)$ nodal region of the Brillouin zone\cite{TYoshida2003,KMShen2004,KMShen2005,TYoshida2006,MHashimoto2008,YYPeng2013,QGao2020}. With increasing doping, a Fermi arc develops around the nodal region in the underdoped regime, characterized by coherent quasiparticles along the arc and incoherent electronic states near the $(\pi,0)$ antinodal region. As the doping further increases, the Fermi arc expands and eventually evolves into a large Fermi surface in the overdoped regime. Such a doping evolution of electronic structure is generally observed in cuprate superconductors and the electronic states around $(\pi,0)$ antinodal region are considered to be particularly important in achieving high temperature superconductivity\cite{JCCampuzano1999,DLFeng2000,HDing2001,XJZhou2004,KMShen2005,KTanaka2006,TDahm2009,ArAbanov2008,DJScalapino2012}.

    In this paper, we report the observation of a unique Fermi surface topology in a seven-layer cuprate Bi$_{2}$Sr$_{2}$Ca$_{6}$Cu$_{7}$O$_{18+\delta}$ (Bi2267). The observed Fermi surface consists of multiple hole-like Fermi pockets centered at $(\pi/2,\pi/2)$, exhibiting distinct momentum-, temperature- and Fermi surface-dependent energy gaps. This system, comprising multiple hole pockets, reveals the coexistence of high temperature superconductivity, antiferromagnetic order and the pseudogap state. These findings challenge the well-accepted picture of doping the parent Mott insulators and provide vital insights into the microscopic origin of high temperature superconductivity in cuprate superconductors.

 \vspace{3mm}

\noindent{\bf Results}

    Our angle-resolved photoemission spectroscopy (ARPES) measurements were conducted on a slightly overdoped Bi$_{2}$Sr$_{2}$Ca$_{2}$Cu$_{3}$O$_{10+\delta}$ (Bi2223) sample with a superconducting transition temperature $T_{\text{c}} = 109.0$\,K (see Methods and Supplementary Fig. S1). It has been shown from the scanning transmission electron microscopy (STEM) measurements that the Bi2223 samples may contain intergrowth phases of Bi$_{2}$Sr$_{2}$Ca$_{n-1}$Cu$_{n}$O$_{2n+4+\delta}$ with $n = 1-9$ (abbreviated as Bi22($n-1$)$n$)\cite{ZWang2023}. Using high-resolution laser-based spatially-resolved ARPES, we successfully identified the multilayer regions on the Bi2223 sample surface (see Methods and Supplementary Fig. S2). The high spatial and energy resolutions of our ARPES setup, combined with excellent system stability, make it possible to systematically collect high quality ARPES data from a fixed spot. While scanning the sample surface, we detected two regions consisting of the seven-layer Bi2267 phase and performed detailed ARPES measurements as reported below.

    Figure 1 presents the Fermi surface mappings and constant energy contours measured in the two regions (Region1 and Region2) of the Bi2223 sample surface. The corresponding band structures along different momentum cuts for Region2 are shown in Fig. 2 (the band structures for Region1 are provided in Supplementary Fig. S3). In both regions, four main Fermi surface sheets are clearly observed and labeled as $\alpha$, $\beta$, $\gamma$ and $\delta$ in Fig. 1(b,e). The Fermi surface mapping for Region2 (Fig. 1e) and the corresponding band structures (Fig. 2) reveal a pocket-like Fermi surface for the $\delta$ band, with weak spectral weight extending across the $(\pi,0)$-$(0,\pi)$ line. The spectral weight of the $\beta$ and $\gamma$ Fermi surfaces is primarily concentrated near the nodal region which can be attributed to a significant anisotropic gap opening, as discussed below. With increasing binding energy, the spectral weight gradually spreads towards the $(\pi,0)$-$(0,\pi)$ line, as seen in the constant energy contours (Fig. 1(c,d,f,g)). The full contours of the $\beta$ and $\gamma$ Fermi pockets become visible (Fig. 1d) and extend smoothly across the $(\pi,0)$-$(0,\pi)$ line. To determine the complete shape of the $\alpha$ Fermi surface, measurements were performed at 80\,K (Fig. 1(h-j)), utilizing thermal broadening to probe the spectral weight above the Fermi level. The $\alpha$ Fermi surface is more consistent with a hole pocket around $(\pi/2, \pi/2)$ based on the analyses of the $\alpha$ band structures and the associated constant energy contours, as shown in Supplementary Figs. S4 and S5, and described in Section 4 of Supplementary Information. We note that, due to the experimentally limited momentum range, the exact topology of the $\alpha$ Fermi surface is difficult to be pinned down and further work needs to be done to expand the covered momentum space up to the antinodal region. The possibility of strong AFM correlations, in addition to AFM order, also needs to be further explored as the origin of the hole pocket-like behavior. By analyzing the Fermi surface mappings and the constant energy contours in Fig. 1(b-j), along with the band structures shown in Fig. 2 and Supplementary Fig. S3, the multiple Fermi pockets are determined as shown in Fig. 1k. These pockets exhibit a similar elliptical shape, with a major-to-minor axis ratio of $1.7 \pm 0.2$. From the areas of these pockets, the estimated doping levels are 0.124 (0.146), 0.076 (0.084), 0.044 (0.055) and 0.024 (0.024) for the $\alpha$, $\beta$, $\gamma$ and $\delta$ Fermi pockets in Region1 (Region2), respectively. Overall, the electronic structures of Region1 and Region2 are similar, with only slight variations in the doping levels of the observed Fermi pockets. We note that, due to the low photon energy of our laser, we cannot directly measure the electronic states near the $(\pi,0)$ and $(0,\pi)$ antinodal regions. No low energy electronic states are observed near the antinodal region from the previous ARPES measurements on the five-layer and six-layer cuprates\cite{SKunisada2020,KKurokawa2023}.

    Figure 1l shows the nodal Fermi momentum $k_{\mathrm{F}}$ of the observed Fermi pockets as a function of their respective doping levels (red symbols in Fig. 1l). Here, $k_{\mathrm{F}}$ is defined as the distance between the $\Gamma$ point and the left edge of the Fermi pocket along the nodal direction (upper-left inset, Fig. 1l). For comparison, we also include data from previous measurements of five-layer and six-layer cuprates\cite{SKunisada2020,KKurokawa2023}. These results show that the nodal Fermi momentum $k_{\mathrm{F}}$ of Fermi pockets decreases monotonically with the increase of the doping level. Additionally, we summarize the distribution of $k_{\mathrm{F}}$ for Bi-based cuprates (Bi2201 and Bi2212) as a function of their doping levels, where a large Fermi surface is typically observed (bottom-right inset, Fig. 1l)\cite{TKondo2007,TKondo2009,YXZhang2016,IKDrozdov2018,YHe2018,YGZhong2018,YDing2019,PAi2019,TValla2020,TValla2021}. Across a broad doping range ($p = 0.06\text{--}0.44$), the nodal Fermi momentum of the large Fermi surface decreases monotonically as the doping level increases. The nodal Fermi momentum has a direct one-to-one correspondence to the doping level of the individual CuO$_{2}$ plane because the interlayer coupling has minimal effect on the nodal Fermi momentum\cite{RSMarkiewicz2005}. As seen in Fig. 1l, the Fermi pocket and the large Fermi surface show distinct doping dependence of the nodal Fermi momentum. It highlights the unique characteristics of the Fermi surface (Fermi pockets) we observed in Bi2267 which are distinct from the large Fermi surface observed in Bi2201 and Bi2212.

    The multiple observed Fermi pockets can be mainly attributed to the four inequivalent types of CuO$_{2}$ planes in the seven-layer cuprate. It has been shown that, in multilayer cuprates, the hole doping level varies significantly among the CuO$_{2}$ planes: it is highest in the outer planes (OPs) due to their proximity to the charge reservoir, decreases progressively toward the inner layers and reaches its lowest value in the innermost plane(s) (IP$_{0}$s)\cite{HMukuda2012,ZWang2023}. In the underdoped or optimally-doped cuprates, the interlayer coupling is usually weak\cite{MMori2006}. In such a case, the formation of a Fermi surface is dominated by a particular type of CuO$_{2}$ plane with a given doping level because the weak interlayer coupling has little effect on the Fermi surface topology\cite{KKurokawa2023}. The observation of separate Fermi pockets indicates that there are four types of CuO$_{2}$ planes in the measured material and the $\alpha$, $\beta$, $\gamma$ and $\delta$ Fermi pockets are mainly derived from the OP, IP$_{2}$, IP$_{1}$ and IP$_{0}$ CuO$_{2}$ planes, respectively. We find that the doping level ratios between adjacent CuO$_{2}$ planes are close to 2, with $p(\mathrm{OP})/p(\mathrm{IP}_{2}) = 1.63$, $p(\mathrm{IP}_{2})/p(\mathrm{IP}_{1}) = 1.73$ and $p(\mathrm{IP}_{1})/p(\mathrm{IP}_{0}) = 1.83$. This trend is consistent with previous observations in multilayer cuprates with an odd number of CuO$_{2}$ planes, which feature a single innermost CuO$_{2}$ plane\cite{SKunisada2020,KKurokawa2023,HChen2026}. In contrast, in multilayer cuprates with an even number of CuO$_{2}$ planes where two innermost CuO$_{2}$ planes (IP$_{0}$s) are present, the doping level of each IP$_{0}$ is approximately halved, yielding a ratio of $p(\mathrm{IP}_{1})/p(\mathrm{IP}_{0}) \approx 4$ in the six-layer system\cite{KKurokawa2023,HChen2026}. Our observed doping distribution among the CuO$_{2}$ planes aligns with the case of odd number of CuO$_{2}$ planes and the region we measured can be attributed to the seven-layer Bi2267, as shown in Fig. 1a.

    Figure 2a illustrates the evolution of the band structures as the momentum cuts rotate from the nodal direction (Cut 1) toward the $(\pi,0)$-$(0,\pi)$ direction (Cut 10) and beyond (Cut 11-Cut 14) around the $(\pi/2,\pi/2)$ point. To enhance the visibility of the bands, the corresponding MDC and EDC second-derivative images are shown in Fig. 2b and Fig. 2c, respectively. Fig. 2e presents typical band structures along momentum cuts nearly parallel to (0,0)-$(\pi,\pi)$ direction around the major vertices of the Fermi pockets. Four distinct bands, labeled $\alpha$, $\beta$, $\gamma$ and $\delta$, are clearly identified and highlighted with colored arrows in Fig. 2(a-c) and Fig. 2e. Along the nodal direction (first panel of Fig. 2(a-c)), the four bands are well separated, with the $\beta$ and $\gamma$ positioned closer to each other. The $\alpha$ band is strong in Fig. 2(a-c) (Cut 1-Cut 9) and Fig. 2e (Cut 15-Cut 17), shifting to higher binding energy from Cut 1 to Cut 9 due to energy gap opening. The $\beta$ band is prominent near the nodal region (Cut 1-Cut 4 in Fig. 2(a-c)) but rapidly loses spectral weight as the momentum cuts rotate from Cut 4 to Cut 10, also shifting to higher binding energy due to energy gap opening. Notably, the $\beta$ band exhibits strong Bogoliubov hybridization with the $\alpha$ back-bending band, which becomes increasingly obvious from Cut 1 to Cut 10. This hybridization, similar to that observed in Bi2223\cite{SKunisada2017,SIdeta2021,XYLuo2023}, likely arises from relatively strong interlayer coupling between OP and IP$_{2}$ layers as well as the substantial superconducting gap difference between the $\alpha$ and $\beta$ bands. The $\gamma$ band remains robust and well-defined from Cut 1 to Cut 10, shifting far below the Fermi level due to the large energy gap opening. It also exhibits Bogoliubov band hybridization with the $\beta$ back-bending band, particularly evident in Cut 7-Cut 9 in Fig. 2(a-c). The $\delta$ band exhibits notably weaker spectral weight compared to the other three bands, particularly at deep energy levels near the $(\pi,0)$-$(0,\pi)$ direction. It crosses the Fermi level all the way from Cut 1 to Cut 13 in Fig. 2(a-c).

    To determine the superconducting gap along the Fermi pockets, the original energy distribution curves (EDCs) and the corresponding symmetrized EDCs along the pockets measured at 20\,K for Region2 are presented in Fig. 3(a-d) and Fig. 3(e-h), respectively. All the four bands show well-defined quasiparticle peaks near the nodal region ($\theta = 0$) (Fig. 3(a-d)). Strong and clear EDC peaks for the $\alpha$ and $\gamma$ bands are observed along the Fermi pockets (Fig. 3(a,c)). The $\beta$ band, being relatively weaker than the $\gamma$ band and overlapping with it, exhibits a two-peak feature in both the EDCs and symmetrized EDCs, as marked in Fig. 3(b,f). The EDC peaks near the Fermi level, marked by ticks in Fig. 3(b,f), correspond to the $\beta$ band, while the higher binding energy peaks originate from the $\gamma$ band. For the $\delta$ band, prominent EDC peaks appear near the nodal region ($\theta = 0$) but become rather weak when the Fermi momentum moves towards the major vertex region ($\theta = 90$) (Fig. 3(d,h)). Detailed analysis of the band structure and EDC feature assignment are summarized in Supplementary Fig. S7. The gap size is extracted from the peak positions in the symmetrized EDCs. The superconducting gap along the $\alpha$, $\beta$, $\gamma$ and $\delta$ Fermi pockets, as obtained from the symmetrized EDCs in Fig. 3(e-h), is plotted in Fig. 3i. Fig. 3j shows the three-dimensional distribution of the energy gaps along the Fermi pockets, revealing distinct Fermi surface-dependent gap behaviours. While the $\alpha$, $\beta$ and $\gamma$ Fermi pockets exhibit clear gap opening, the $\delta$ Fermi pocket remains gapless along the Fermi pocket.

    The gap opening along the $\alpha$, $\beta$ and $\gamma$ Fermi pockets is highly anisotropic, vanishing along the nodal direction ($\theta = 0$) and reaching its maximum near the major vertex ($\theta = 90$). The momentum-dependence of the gap for the $\alpha$ and $\beta$ Fermi pockets follows a d-wave form $\Delta=\Delta_{0}|\cos(k_{x}a)-\cos(k_{y}a)|/2$ as seen in Fig. 3k. The gap along the $\gamma$ Fermi pocket follows the d-wave form near the nodal region ($\theta = 0\sim70$) but deviates near the major vertex region ($\theta = 70\sim90$). This gap behaviour is similar to that observed in underdoped cuprates\cite{MHashimoto2014}. The gap sizes along the $\alpha$, $\beta$ and $\gamma$ Fermi pockets differ significantly. The maximum gap sizes at the major vertex are $\sim$19, $\sim$32.5 and $\sim$42\,meV for the $\alpha$, $\beta$ and $\gamma$ Fermi pockets, respectively (Fig. 3(i-k)). The extrapolated maximum gap sizes $\Delta_{\mathrm{0}}$, extracted from the d-wave form in Fig. 3k, are $\sim$29, $\sim$65 and $\sim$76\,meV for the $\alpha$, $\beta$ and $\gamma$ Fermi pockets, respectively. The EDC lineshape, momentum-dependence and Fermi surface-dependence of the energy gap along the Fermi pockets are similar from the measurements of Region1, except for slight variations in gap size due to doping differences between the two measured regions (Supplementary Fig. S8). Fig. 3l summarizes the maximum energy gap $\Delta_{\mathrm{max}}$ (purple symbols) and the extrapolated energy gap $\Delta_{\mathrm{0}}$ (black symbols) along the Fermi pockets measured in both Region1 and Region2. We note that the energy gaps along the $\alpha$, $\beta$ and $\gamma$ Fermi pockets we observed in our seven-layer Bi2267 are significantly larger than those reported for five-layer and six-layer cuprates\cite{SKunisada2020,KKurokawa2023}. In the previous ARPES measurements on the five-layer and six-layer cuprates\cite{SKunisada2020,KKurokawa2023}, only a small gap ($< 5$\,meV) is observed in the IP$_1$ CuO$_2$ plane adjacent to OP which is comparable or slightly smaller than the gap in the OP CuO$_2$ plane. This suggests that the superconducting gap in the IP$_1$ CuO$_2$ plane is possibly induced by the OP CuO$_2$ plane. In our seven-layer cuprate, (1). we observed much larger gap opening (up to 42\,meV) in the OP, IP$_2$ and IP$_1$ CuO$_2$ planes; (2). we observed a large gap opening in the IP$_1$ CuO$_2$ plane that is not adjacent to OP CuO$_2$ plane. Its gap size is even larger than that in the IP$_2$ CuO$_2$ plane that is adjacent to the OP CuO$_2$ plane; (3). The gap opening in the inner IP$_1$ and IP$_2$ CuO$_2$ planes is significantly larger than that in the OP CuO$_2$ plane. These results clearly indicate that the gap opening in the inner CuO$_2$ planes is not induced by the outer OP CuO$_2$ planes. Instead, the gap opening is more intrinsic to a specific CuO$_2$ plane which is dictated by electron correlation and magnetic interaction that are dependent on the doping level and the interlayer interactions. The extrapolated energy gap $\Delta_{0}$, derived from energy gaps around the nodal region (Fig. 3k), is generally regarded as the intrinsic superconducting gap. For the $\gamma$ Fermi pocket, this gap can reach $\sim$76\,meV (Fig. 3l) exceeding that of Bi2223 and representing the largest value reported in any cuprates to date\cite{MHashimoto2014,SIdeta2010}. The large values of $\Delta_{0,\beta}$ and $\Delta_{0,\gamma}$ (Fig. 3l) indicate the strong pairing strength in the $\beta$ and $\gamma$ Fermi pockets of Bi2267.

    To investigate the temperature dependence of the gap opening and superconductivity in the Fermi pockets, we performed detailed measurements in Region2 over a temperature range of 20\,K to 115\,K (Fig. 4). Fig. 4a shows a typical band structure measured at different temperatures along a momentum cut near the major vertices of the Fermi pockets ($\theta = 50$, Fig. 4b). The four bands, $\alpha$, $\beta$, $\gamma$ and $\delta$, are clearly resolved and exhibit distinct temperature evolution. The $\delta$ band crosses the Fermi level without forming a gap at all measured temperatures. Fig. 4(c-e) display the EDCs at different temperatures measured at the Fermi momenta of the $\alpha$, $\beta$ and $\gamma$ bands, respectively, obtained from Fig. 4a. The corresponding symmetrized EDCs are presented in Fig. 4(f-h). Further analyses are shown in Supplementary Figs. S9 and S10. At low temperatures, strong and sharp EDC peaks appear for the $\alpha$ (Fig. 4c) and $\gamma$ (Fig. 4e) bands, whereas the $\beta$ band sharpens but does not form well-defined peaks (Fig. 4d).

    The $\alpha$, $\beta$ and $\gamma$ Fermi pockets exhibit distinct temperature-dependent behaviors in the energy gap opening and superconducting coherence development. As shown from Fig. 4(f-h), the $\alpha$ band develops an energy gap at low temperatures but the gap closes at 80\,K and 100\,K (Fig. 4f). In contrast, the gap opening in the $\beta$ and $\gamma$ bands persists up to 100\,K (Fig. 4(g,h)). Fig. 4i presents the temperature dependence of the spectral intensity of the dip at the Fermi level, the EDC peak height and their difference obtained from the symmetrized EDCs for the $\alpha$ band in Fig. 4f. The spectral weight shows little variation with temperature between 80\,K and 115\,K, while the superconducting coherence peak starts to abruptly develop between 70\,K and 80\,K. This provides a reliable estimate of the superconducting transition temperature ($T_{\mathrm{c}}$) of the measured Region2 as $75\pm5$\,K\cite{YXu2021,SDChen2022}. This is consistent with the $T_{\mathrm{c}}$ of $70\pm5$\,K for the seven-layer Bi2267 reported in Refs.\cite{ZWang2023,HNarita1992} and the overall gap measurements along the three Fermi pockets at different temperatures (Fig. 4(j-l)). For the $\alpha$ Fermi surface, below 70\,K, the energy gap is present along the entire Fermi pocket except for the nodal direction (Fig. 4j). At 80\,K and 90\,K, the gap closes around the nodal region, forming a Fermi arc extending to $\theta = 50$. At 100\,K, the gap vanishes along the entire Fermi pocket. These observations suggest a superconducting transition temperature $T_{\mathrm{c}}$ between 70-80\,K, the emergence of a pseudogap between 70-100\,K and the pseudogap temperature onset $T^*$ of $95\pm5$\,K for the $\alpha$ Fermi surface. For the $\beta$ and $\gamma$ Fermi pockets, a well-defined Fermi arc appears at 100\,K, with a length extending only to $\theta = 30$ (Fig. 4(k,l)). This indicates that the pseudogap temperature $T^*$ of the $\beta$ and $\gamma$ Fermi pockets is significantly higher than that of the $\alpha$ Fermi surface. Such a variation in $T^*$ can be attributed to the doping difference of the Fermi pockets with the $\alpha$ Fermi surface being slightly underdoped ($p = 0.146$), while the $\beta$ ($p = 0.084$) and $\gamma$ ($p = 0.055$) pockets are heavily underdoped. Notably, this represents a direct observation of pseudogap formation on the Fermi pockets by ARPES in multilayer cuprates.

    It has been found that the antiferromagnetic order is strongly enhanced in multilayer cuprates and the strength of the antiferromagnetic order increases with the number of CuO$_{2}$ layers ($n$)\cite{HMukuda2012}. Our observation of multiple Fermi pockets is consistent with the presence of strong antiferromagnetic order and strong electron correlation in this seven-layer cuprate. The observation that the $\alpha$ Fermi surface is more consistent with the Fermi pocket picture suggests that there remain strong electron correlation and antiferromagnetic order in the outer CuO$_2$ planes in the seven-layer cuprates. As shown in \cite{HMukuda2012}, for the five-layer ($n = 5$) cuprates, the OP layers can be antiferromagnetic in the underdoped sample (see Fig. 6a of \cite{HMukuda2012}). Since the region of the long-range antiferromagnetic order increases with the number of CuO$_2$ planes ($n$), it is reasonable to expect that the OP layers in the seven-layer ($n = 7$) cuprates may exhibit a static antiferromagnetic order when the doping level is relatively low. The Fermi pockets we observed in Bi2267 are different from the pockets observed in the electron doped cuprates \cite{HMatsui2005,JHe2019}. Their formation cannot be explained in the band folding picture. When the doping level of CuO$_2$ plane is small, there is no large Fermi surface to start with as required in the band folding picture. Instead, the electron correlation is strong and it is more suitable to be described by the $t$-$U$ model as presented in detail in the Supplementary Information\cite{JRSchrieffer1989,AVChubukov1992,NBulut1994,TXiang1996,CKusko2002}. In this framework, key parameters such as the intralayer electron hopping ($t$, $t^{'}$ and $t^{''}$), the chemical potential ($\mu$), the mean field gap ($\Delta_{\mathrm{MF}}$) and the superconducting gap ($\Delta_{\mathrm{SC}}$) for each layer, along with the interlayer hopping terms ($t_{\mathrm{oi2}}$, $t_{\mathrm{i1i2}}$ and $t_{\mathrm{i0i1}}$), are taken into account (Fig. 5a). The observed Fermi surface (Fig. 2d) and the measured band structures along typical momentum cuts in the normal state (Fig. 5d and Supplementary Fig. S11) for Region2 are well fitted by using the mean field $t$-$U$ model, with the fitted parameters plotted in Fig. 5(g-k). The simulated Fermi surface (Fig. 5b) and the corresponding simulated band structures in the normal state (Fig. 5c and Supplementary Fig. S11) obtained by using these parameters in Fig. 5(g-k), are in good agreement with the measured data. Incorporating a set of appropriate superconducting gap and interlayer coupling parameters (listed in the Supplementary Information and Supplementary Table S1), we further simulated the band structures in the superconducting state (Fig. 5e). These simulations also closely match with the measured band structures in Fig. 5f.

    The observation of multiple Fermi pockets demonstrates that strong electron correlation and strong antiferromagnetic order are associated with all the seven CuO$_{2}$ planes in Bi2267. From the fitted parameters (Fig. 5(g-k)), we find that the multiple Fermi pockets and their associated band structures in the normal state can be described by a similar set of parameters $\{t, t^{\prime}, t^{\prime\prime}, \Delta_{\mathrm{MF}}\} = \{0.4, -0.079, 0.032, 0.75\}$ and a varying $\mu$. Based on the same band structure (Supplementary Fig. S11b), the formation of the multiple Fermi pockets can be attributed to the shifting of the chemical potential (Supplementary Fig. S11(a,b)). The variation in chemical potential arises from the uneven charge carrier distribution among the four types of CuO$_{2}$ planes in Bi2267 (Fig. 5a). Our simulations further indicate that the interlayer coupling has a minimal impact  on the Fermi pocket topology and related band structures (Supplementary Fig. S12). As the interlayer coupling strength depends on the doping levels of adjacent CuO$_{2}$ planes\cite{MMori2006}, and the three inner planes are underdoped or heavily underdoped (Fig. 5a), the interlayer coupling is weak in Bi2267, particularly between the two inner planes. Consequently, the observed Fermi pockets ($\alpha$, $\beta$, $\gamma$ and $\delta$) in momentum space can be mainly attributed to the four distinct types of CuO$_{2}$ planes (OP, IP$_{2}$, IP$_{1}$ and IP$_{0}$) in real space.

 \vspace{3mm}

 \noindent{\bf Discussion}

    In Bi2267, the doping level of the IP$_1$ (IP$_2$) CuO$_2$ plane is $\sim0.05$ ($\sim0.08$), which is heavily underdoped. We found that the pairing gap along the IP$_1$ (IP$_2$) Fermi pocket can reach $\Delta_{\mathrm{max}} = 42$\,meV ($\sim33$\,meV) and the extrapolated gap $\Delta_{0} = 76$\,meV ($\sim66$\,meV) is even larger than the gap $\Delta_{0}\sim60$\,meV for the inner IP CuO$_2$ plane in Bi2223 that has a doping level of 0.08 (Fig. 5l). The observation of larger gap opening on the IP$_1$ and IP$_2$ CuO$_2$ planes in Bi2267 is consistent with the usual trend that the energy gap size increases with decreasing doping level in cuprates\cite{VJEmery1995}. In cuprate superconductors, the superconducting transition temperature $T_{\mathrm{c}}$ is determined not only by the gap size, but also by the coherence temperature, which increases with increasing doping level\cite{VJEmery1995}. In multilayer cuprates, the pairing strength and the coherence temperature may be realized in different CuO$_2$ planes, and in this case, the interlayer coupling is also important in determining $T_{\mathrm{c}}$. In optimally-doped and overdoped Bi2223, the OP CuO$_2$ planes are heavily overdoped ($p = 0.22\text{--}0.30$)\cite{SIdeta2010,XYLuo2023}. In Bi2267, the doping level of the OP CuO$_2$ planes is $p = 0.12\text{--}0.15$. This indicates that the coherence temperature of the OP CuO$_2$ planes in Bi2267 is lower than that in Bi2223. The coherence temperature difference may account for why the maximum $T_{\mathrm{c}}$ in Bi2223 ($T_{\mathrm{c}}\sim110$\,K) is higher than that in Bi2267 ($T_{\mathrm{c}}\sim75$\,K).

    Our findings provide the direct observation of a cuprate exhibiting multiple hole Fermi pockets and demonstrate that such a system can sustain high temperature superconductivity with a $T_{\mathrm{c}}$ as high as $\sim75$\,K. In previous studies, cuprates have consistently exhibited a large Fermi surface, although the antinodal states may become incoherent in the underdoped regime\cite{TYoshida2003,KMShen2005,UChatterjee2010}. The antinodal electronic states have been considered essential for high temperature superconductivity\cite{DLFeng2000,HDing2001,XJZhou2004,KMShen2005,TDahm2009,ArAbanov2008,DJScalapino2012}. However, our present observation of Fermi pockets only indicates that the nodal region electronic states alone can support high temperature superconductivity, challenging the conventional understanding of the role of antinodal electronic states. Although the observed Fermi pockets are confined near the nodal region, a large energy gap of up to $\sim42$\,meV can already open along the Fermi pockets (Fig. 3l) indicative of strong pairing strength that can be realized in the Fermi pockets. In fact, Bi2267 exhibits the strongest pairing strength among all measured cuprate superconductors (Fig. 5l)\cite{SIdeta2010,IMVishik2012,MHashimoto2014,XYLuo2023}. Furthermore, the $\alpha$ and $\gamma$ bands exhibit pronounced superconducting coherence peaks at low temperatures (Fig. 3 (a,c)), reflecting the establishment of phase coherence in the corresponding CuO$_{2}$ planes. These indicate that the Fermi pockets alone meet the requirements for the strong pairing strength and phase coherence to achieve high temperature superconductivity\cite{VJEmery1995}. Our observation of high temperature superconductivity in a cuprate with multiple hole pockets raises a fundamental question: in cuprates with a large Fermi surface, which part of the electronic states, the nodal region or antinodal region, plays the dominant role in generating high temperature superconductivity. Our present findings motivate further theoretical studies to investigate whether and how high temperature superconductivity can be realized through only the nodal region electronic states.

    Our present work also provides valuable perspectives on the electronic phase diagram and electron pairing mechanism in cuprates. In the conventional phase diagram of cuprates, high temperature superconductivity emerges only after the suppression of antiferromagnetic order\cite{BKeimer2015}, which implies the competition between superconductivity and the antiferromagnetic order. However, in our Bi2267, we observe that high temperature superconductivity can coexist with the strong antiferromagnetic order. In particular, strong electron pairing with an energy gap as high as $\sim42$\,meV can be realized in a CuO$_{2}$ plane (IP$_{1}$) with a doping level of $\sim0.05$ and a robust antiferromagnetic order (Fig. 3l). This observation offers deeper insight to understand the pairing mechanism in cuprate superconductors. The previous NMR studies on multilayer cuprates\cite{HMukuda2012} indicate that the region of the long-range antiferromagnetic order increases with the number of CuO$_2$ planes ($n$). For $n = 5$, long-range AFM order can persist up to a doping level of $\sim 0.10$, and for our case of $n = 7$, such a doping level is expected to be even higher. The slightly doped CuO$_2$ planes in the seven-layer cuprate ($p\sim0.05$) therefore are well within the doping range of the long-range antiferromagnetic order. We found that all the observed $\gamma$ and $\delta$ bands are caused by Fermi pockets; no other features are observed outside of the Fermi pockets. These results also indicate that the IP$_1$ and IP$_0$ layers are in the long-range AFM order regime. The electron pairing already occurs in the slightly doped CuO$_{2}$ planes (p$\sim$0.05) which exhibit an antiferromagnetic order without entering the spin fluctuation region yet\cite{HMukuda2012}. These findings indicate that there is already a pairing interaction in the antiferromagnetic state that is distinct from the spin fluctuations\cite{DJScalapino1986,PMonthoux1991,DJScalapino1995,TMoriya2000,ArAbanov2008,DJScalapino2012}. Further investigations are required to clarify how strong electron pairing emerges in the CuO$_{2}$ plane with slight doping and robust antiferromagnetic order.

    In summary, by using high-resolution laser-based spatially-resolved ARPES, we successfully measured the electronic structure and energy gap of the seven-layer Bi2267. We identified a cuprate system that consists of multiple hole Fermi pockets. The formation of Fermi pockets originates from the strong electron correlation and the enhanced antiferromagnetic order in the multilayer system. Strong electron pairing with an energy gap as high as $\sim42$\,meV is observed in the CuO$_{2}$ plane at a slight doping level of $\sim0.05$ where a robust antiferromagnetic order persists. The large energy gap and sharp EDC peaks observed along the Fermi pockets near the nodal region highlights the strong electron pairing strength and phase coherence. Importantly, such a system with multiple hole pockets can sustain high temperature superconductivity with a $T_{\mathrm{c}}$ of $\sim75$\,K while coexisting with the antiferromagnetic order. These observations provide key insights in understanding the electronic phase diagram and the electron pairing mechanism in high temperature cuprate superconductors.

\vspace{3mm}

\noindent{\bf Methods}

\noindent{\bf Sample growth and preparation}

  Single crystals of Bi$_{2}$Sr$_{2}$Ca$_{2}$Cu$_{3}$O$_{10+\delta}$ (Bi2223) were grown by the traveling solvent floating zone method\cite{BLiang2002,CTLin2002}. The samples were post-annealed at 550\,$^{\circ}$C under high oxygen pressure about 160 atmospheres for 14 days. The obtained samples are slightly overdoped with a $T_{\mathrm{c}}$ of 109.0\,K and a transition width $\Delta$T $\approx$2.0\,K, as seen in Supplementary Fig. S1.

\noindent{\bf ARPES measurements}

  ARPES measurements were carried out using our lab-based laser ARPES system equipped with the 6.994 eV vacuum-ultra-violet (VUV) laser and a DA30L hemispherical electron energy analyzer\cite{XJZhou2008GDLiu,WTZhang2018XJZhou}. The energy resolution was set at 1\,meV and the angular resolution is approximately 0.3$^{\circ}$, corresponding to a momentum resolution of 0.004 $\AA^{-1}$.  The samples were cleaved \textit{in situ} at a low temperature and measured in vacuum with a base pressure better than $5\times10^{-11}$ Torr at 20\,K. The Fermi level is referenced by measuring on a clean polycrystalline gold that is electrically connected to the sample. The laser spot size on the sample was set at $\sim$20\,$\mu$m. It has been observed that our Bi2223 single crystal samples contain intergrowth phases of Bi$_{2}$Sr$_{2}$Ca$_{n-1}$Cu$_{n}$O$_{2n+4+\delta}$ with the number of CuO$_{2}$ planes per structural unit $n = 1-9$\cite{ZWang2023}. We first scanned the entire sample surface by performing ARPES measurements on each spot (see Supplementary Fig. S2). Different regions of the sample surface were identified by looking at the number of bands and their distribution measured on each spot. In this work, we selected the region with four bands (Supplementary Fig. S2h) with a size of $\sim50$\,$\mu$m and carried out detailed momentum- and temperature-dependent measurements. The DA30 mode of our electron energy analyser is used to map the Fermi surface and band structures without moving the sample at different temperatures. From the measured Fermi surface (Fig. 1) and band structures (Fig. 2), we determine that the measured area is Bi2267 as described in the main text.\\

\vspace{3mm}

\noindent {\bf Data availability}\\
All data is processed by using Igor Pro 8.02 software. All data needed to evaluate the conclusions in the paper are available within the article and its Supplementary Information files. All raw data generated during the current study are available from the corresponding author upon request.\\

    \vspace{3mm}

    \noindent {\bf Acknowledgements}\\
This work is supported by the National Natural Science Foundation of China (Grant Nos. 12488201 by X.J.Z., 12374066 by L.Z. and 12374154 by X.T.L.), the National Key Research and Development Program of China (Grant Nos. 2021YFA1401800 by X.J.Z., 2022YFA1604200 by L.Z., 2022YFA1403900 by G.D.L., 2023YFA1406000 and 2024YFA1408300 by X.T.L.), the Strategic Priority Research Program (B) of the Chinese Academy of Sciences (Grant No. XDB25000000 by X.J.Z.), Innovation Program for Quantum Science and Technology (Grant No. 2021ZD0301800 by X.J.Z.), the Youth Innovation Promotion Association of CAS (Grant No. Y2021006 by L.Z.) and the Synergetic Extreme Condition User Facility (SECUF).

    \vspace{3mm}

    \noindent {\bf Author Contributions Statement}\\
    X.J.Z., L.Z. and X.Y.L. proposed and designed the research. X.Y.L., H.C., Y.W.C. and J.M.S. carried out the ARPES experiments. C.T.L. grew the single crystals. X.Y.L. contributed in sample preparation. H.C., T.M.M., B.L., W.P.Z., N.C., X.L.R., Y.J.S., C.H.Y., J.X.Z., S.J.Z., Z.M.W., F.F.Z., F.Y., Q.J.P., Z.Y.X., G.D.L., X.T.L., H.Q.M., L.Z. and X.J.Z. contributed to the development and maintenance of the ARPES systems and related software development. X.Y.L., Y.H.L. and T.X. contributed to theoretical analysis and discussions. X.Y.L., L.Z. and X.J.Z. analyzed the data and wrote the paper. All authors participated in discussions and comments on the paper.

    \vspace{3mm}

    \noindent{\bf Competing Interests Statement}\\
     The authors declare no competing interests.

    \vspace{3mm}
    \noindent {\bf Figure Legends}

    \begin{figure*}[h]
    \begin{center}
    \includegraphics[width=1\columnwidth,angle=0]{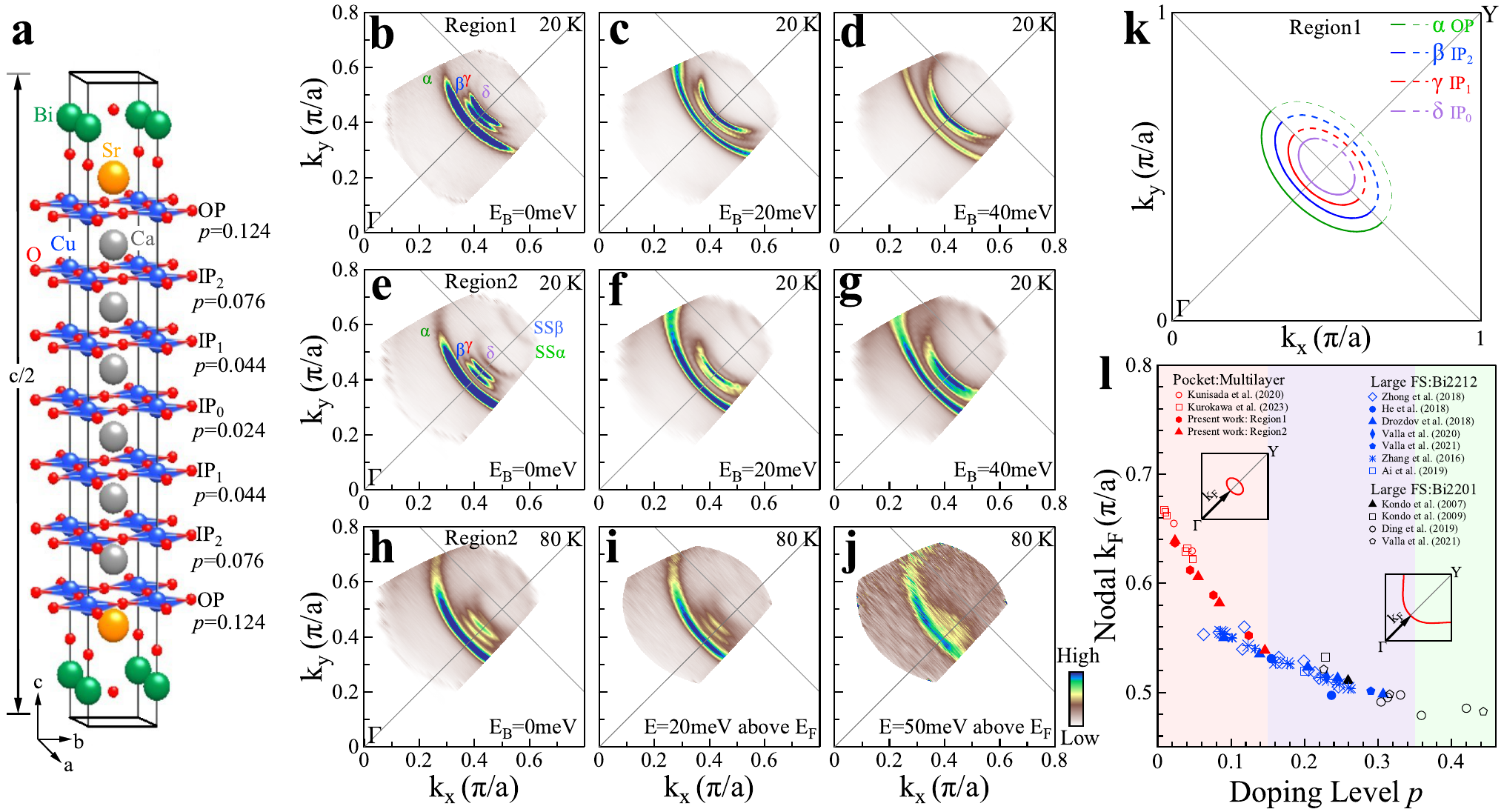}
    \end{center}
    \caption{{\bf Observation of multiple Fermi pockets in Bi2267.} (a) Crystal structure of Bi2267. Each half unit cell contains seven adjacent CuO$_{2}$ planes with three kinds of inner planes (IP$_{0}$, IP$_{1}$, IP$_{2}$) and one kind of outer planes (OP). The doping level progressively decreases from the outermost to the innermost planes, with values of 0.124, 0.076, 0.044 and 0.024 for the OP, IP$_{2}$, IP$_{1}$ and IP$_{0}$ planes, respectively, deduced from the Fermi pocket areas in Region1. (b-d) Fermi surface mapping (b) and constant energy contours at binding energies of 20\,meV (c) and 40\,meV (d) measured at 20\,K in Region1. (e-g) Same measurements as (b-d) but in Region2. (h-j) Fermi surface mapping (h) and constant energy contours at energies of 20 (i) and 50\,meV (j) above the Fermi level ($E_{\mathrm{F}}$) measured at 80\,K in Region2. Four Fermi surface sheets are identified as $\alpha$, $\beta$, $\gamma$ and $\delta$ along with the superstructure Fermi surface denoted as SS$\alpha$ and SS$\beta$. Detailed analysis is shown in Supplementary Fig. S6. (k) Fermi surface of Bi2267 in Region1. Multiple Fermi pockets are observed labelled as $\alpha$, $\beta$, $\gamma$ and $\delta$, which are mainly derived from OP, IP$_{2}$, IP$_{1}$ and IP$_{0}$ planes, respectively. On the opposite side of the ($\pi$,0)-(0,$\pi$) line, except for part of the $\delta$ Fermi pocket, the observed Fermi pockets are rather weak and are plotted as dashed lines. (l) Summarized doping dependence of the nodal Fermi momentum ($k_{\mathrm{F}}$) in some cuprate superconductors. The nodal Fermi momentum is defined in the upper-left inset for a Fermi pocket and in the bottom-right inset for a large Fermi surface. The data points combine our results with those from the previous measurements of Bi2201, Bi2212 and other multilayer cuprates\cite{SKunisada2020,KKurokawa2023,TKondo2007,TKondo2009,YXZhang2016,IKDrozdov2018,YHe2018,YGZhong2018,YDing2019,PAi2019,TValla2020,TValla2021}.}

\end{figure*}

\clearpage

\begin{figure*}[tp]
\begin{center}
\includegraphics[width=1.0\columnwidth,angle=0]{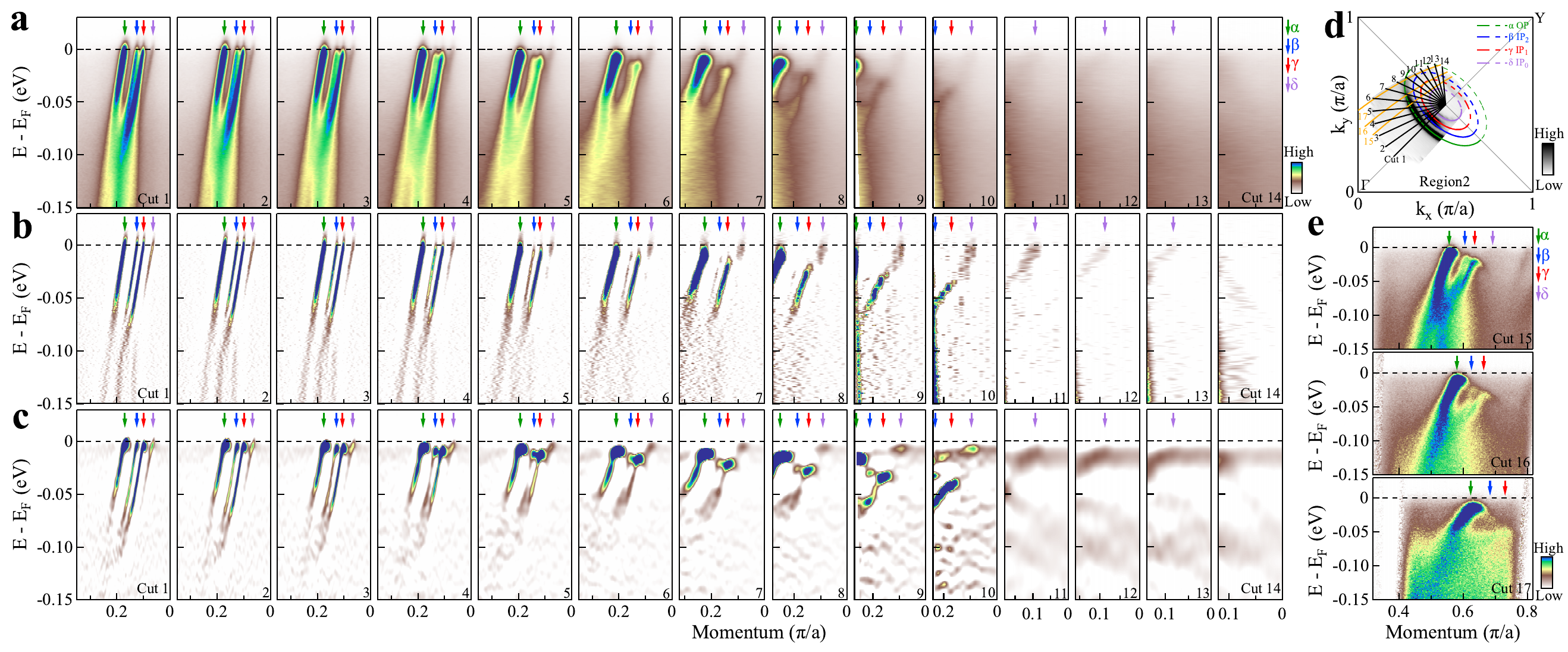}
\end{center}
	 \caption{{\bf Momentum-dependent band structures of Bi2267 measured at 20\,K in Region2.} (a) Band structures measured along various momentum cuts crossing $(\pi/2,\pi/2)$ point. The locations of these momentum cuts are indicated by black solid lines in (d). Four primary bands are identified and labeled as $\alpha$, $\beta$, $\gamma$ and $\delta$ marked by green, blue, red and lilac arrows, respectively. (b) The corresponding MDC second derivative images from (a). (c) The corresponding EDC second derivative images from (a). (d) Fermi surface mapping and the multiple Fermi pockets $\alpha$ (green line), $\beta$ (blue line), $\gamma$ (red line) and $\delta$ (lilac line), with the momentum cuts marked. (e) Band structures measured along the momentum Cuts 15-17; the location of the momentum cuts is shown in (d) as orange solid lines. The four primary bands $\alpha$, $\beta$, $\gamma$ and $\delta$ are marked by green, blue, red and lilac arrows, respectively.}

\end{figure*}

\clearpage

\begin{figure*}[tp]
\begin{center}
\includegraphics[width=1.0\columnwidth,angle=0]{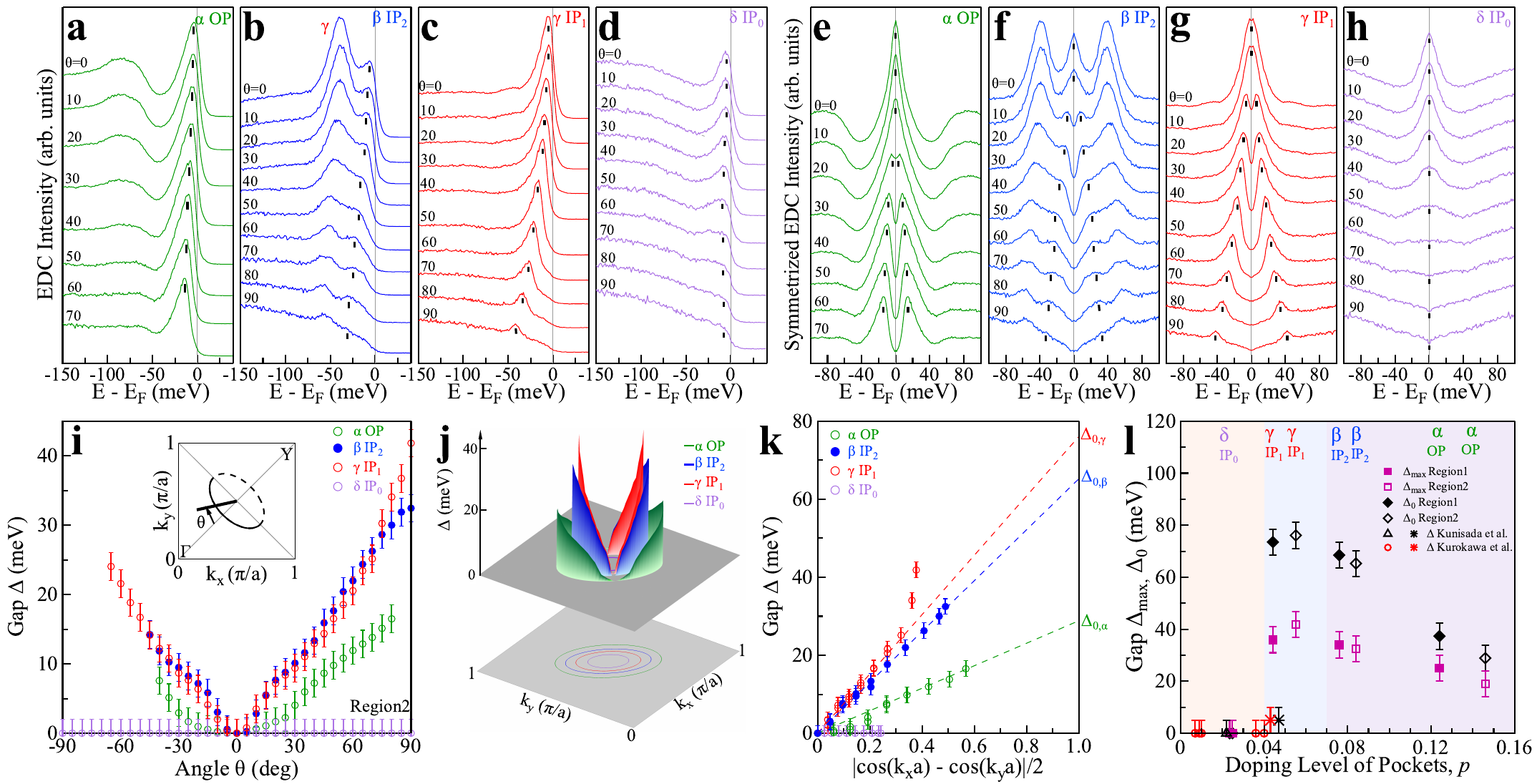}
\end{center}
	\caption{{\bf Photoemission spectra and the energy gap of Bi2267 along the Fermi pockets measured at 20\,K in Region2.} (a-d) EDCs measured along the $\alpha$ (a), $\beta$ (b), $\gamma$ (c) and $\delta$ (d) Fermi pockets, respectively. The location of the Fermi momentum is defined by the angle $\theta$, as depicted in the inset of (i). The EDC peaks corresponding to $\alpha$, $\beta$, $\gamma$ and $\delta$ bands are marked by ticks in (a-d). (e-h) Symmetrized EDCs obtained from (a-d). (i) Momentum-dependent energy gaps along the Fermi pockets obtained from the symmetrized EDCs in (e-h). The maximum gap size at the major vertex is $\sim19$\,meV, $\sim32.5$\,meV, $\sim42$\,meV and 0 for the $\alpha$, $\beta$, $\gamma$ and $\delta$ Fermi pockets, respectively. The angle $\theta$ is defined in the insets of (i). (j) Three-dimensional plot of the energy gaps along the Fermi pockets. (k) The energy gaps plotted as a function of $|\cos(k_{x}a)-\cos(k_{y}a)|/2$ for the Fermi pockets. The energy gaps along $\alpha$ and $\beta$ Fermi pockets exhibit a linear dependence, with the extrapolated values $\Delta_{0,\alpha}$ and $\Delta_{0,\beta}$ at the $(0,\pi)$ point. The energy gap along $\gamma$ Fermi pocket shows a linear dependence near the nodal region with the extrapolated value $\Delta_{0,\gamma}$ at the $(0,\pi)$ point. It deviates from a linear dependence with the momentum moving away from the nodal region. (l) The distribution of the energy gaps, the maximum energy gap $\Delta_{\mathrm{max}}$ (purple symbols) and the extrapolated energy gap $\Delta_{0}$ (black symbols), on the Fermi pockets measured in two regions (Region1 and Region2). Error bars indicate the uncertainties in determining the energy gap sizes. The maximum energy gaps observed in the previous work\cite{SKunisada2020,KKurokawa2023} are also included for comparison.
}

\end{figure*}

\clearpage

\begin{figure*}[tp]
\begin{center}
\includegraphics[width=1.0\columnwidth,angle=0]{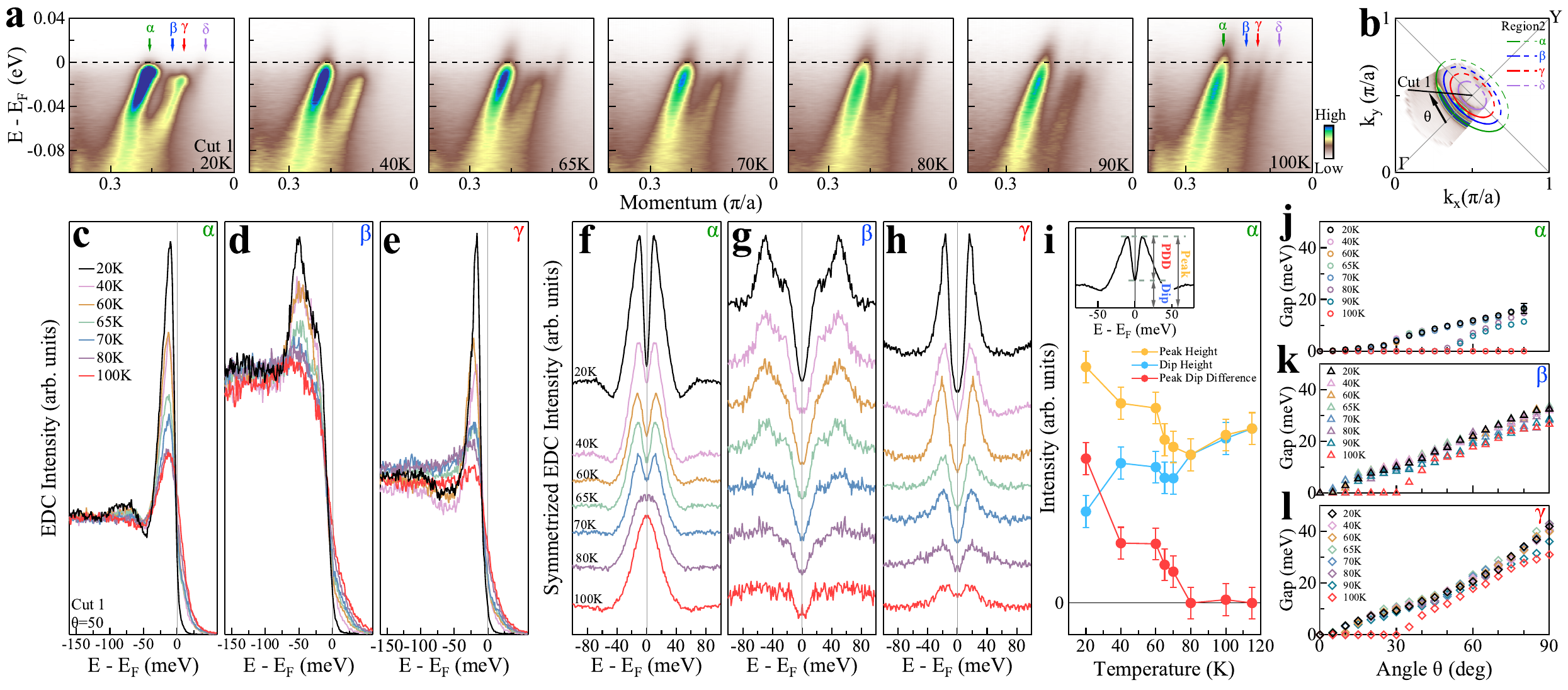}
\end{center}
	\caption{{\bf Evolution of band structures, photoemission spectra and energy gaps with temperature along a typical momentum cut crossing the Fermi pockets.} (a) Band structures measured along the momentum Cut 1 at different temperatures. The location of the momentum Cut 1 is shown in (b). (b) Fermi surface mapping and the Fermi pockets $\alpha$ (green line), $\beta$ (blue line), $\gamma$ (red line) and $\delta$ (lilac line), with the momentum cut marked and the angle $\theta$ defined. (c-e) EDCs measured at different temperatures obtained from (a) at the Fermi momentum of the $\alpha$ (c), $\beta$ (d) and $\gamma$ (e) bands as marked by the colored arrows in (a). (f-h) The corresponding symmetrized EDCs obtained from (c-e). (i) Temperature dependence of the spectral intensities extracted from the symmetrized EDCs of the $\alpha$ band in (f). The upper-left inset defines the three spectral intensities: the peak height, the dip height and the peak-dip difference (PDD). The corresponding intensities as a function of temperature are plotted by the orange solid circles (peak height), the blue solid circles (dip height) and the red solid circles (PDD). The error bars reflect uncertainties in determining the spectral intensity. (j-l) Momentum dependence of the energy gap along the $\alpha$ (j), $\beta$ (k) and $\gamma$ (l) Fermi pockets measured at different temperatures. The uncertainties in determining the gap size are marked by the error bar in (j).}

	\end{figure*}
\clearpage

\begin{figure*}[ht]
\begin{center}
\includegraphics[width=1.0\columnwidth,angle=0]{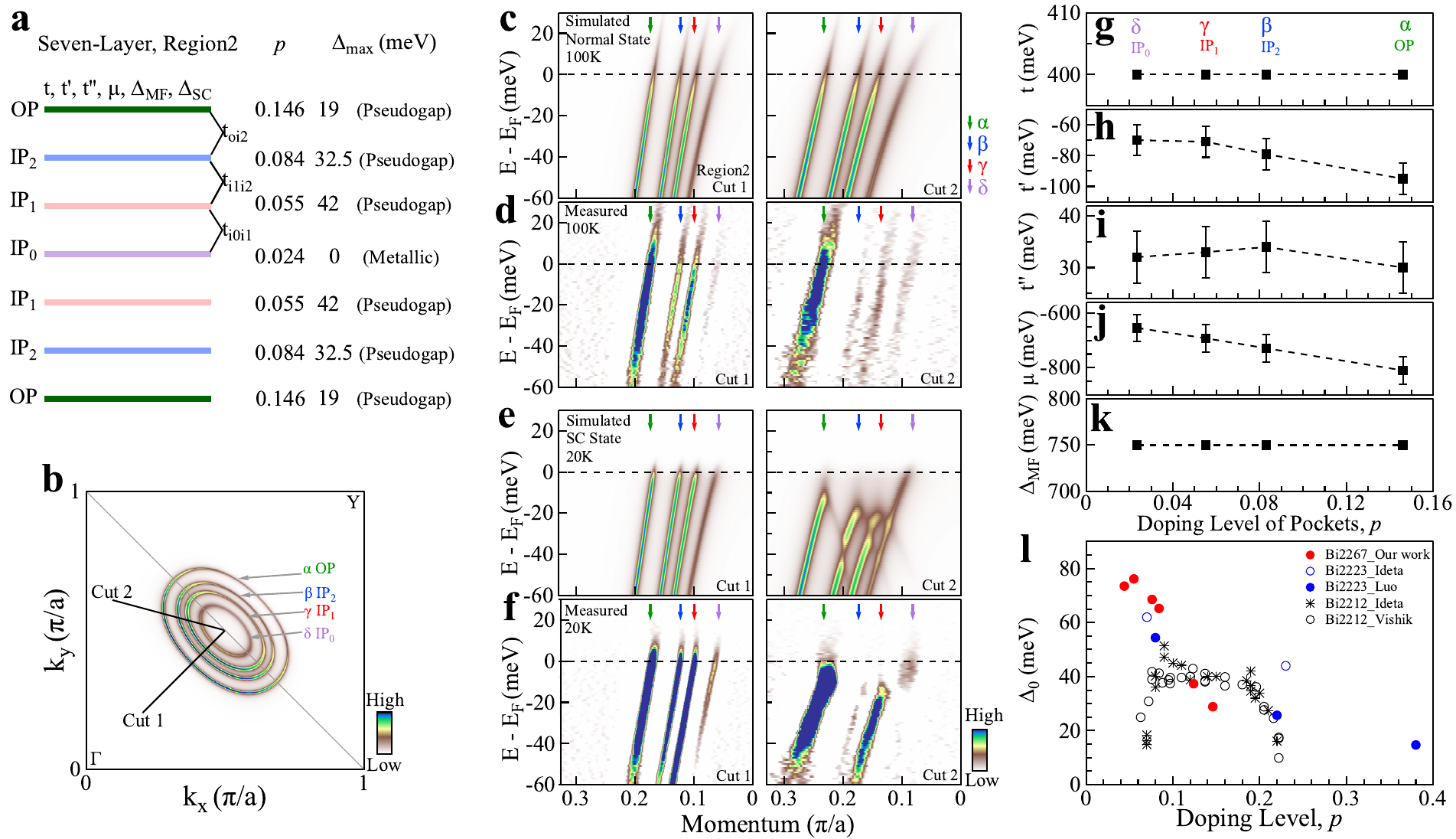}
\end{center}
	\caption{{\bf Formation of the multiple Fermi pockets in the seven-layer cuprate in terms of the mean field $t$-$U$ model.} (a) Schematic illustration of the seven-layer mean field $t$-$U$ model including the in-plane parameters $t$, $t^{\prime}$, $t^{\prime\prime}$, $\mu$, $\Delta_{\mathrm{MF}}$ and $\Delta_{\mathrm{SC}}$, and the interlayer hopping parameters $t_{\mathrm{oi2}}$, $t_{\mathrm{i1i2}}$ and $t_{\mathrm{i0i1}}$. The doping level and maximum energy gap distributions among the seven CuO$_{2}$ layers in Bi2267 (Region2) are also shown. (b) Simulated Fermi surface consisting of multiple Fermi pockets. (c-f) Simulated band structures (c,e) and their comparison with the measured results (d,f). The simulation details are described in Section 9 of the Supplementary Information. (c) Simulated band structures along the momentum Cut 1 (left panel) and Cut 2 (right panel) at 100\,K in the normal state. The location of the momentum cuts is represented by black solid lines in (b). Four bands are labeled as $\alpha$, $\beta$, $\gamma$ and $\delta$ marked by green, blue, red and lilac arrows, respectively. (d) Measured band structures along the momentum Cut 1 (left panel) and Cut 2 (right panel) at 100\,K in the normal state in Region2. These are MDC second-derivative images. (e-f) Same as (c-d) but at 20\,K in the superconducting state. (g-k) The extracted parameters $t$ (g), $t^{\prime}$ (h), $t^{\prime\prime}$ (i), $\mu$ (j) and $\Delta_{\mathrm{MF}}$ (k) obtained by fitting the observed Fermi surface and band structures in Region2 with the mean field $t$-$U$ model. Error bars in (g-k) represent the uncertainties in extracting the fitting parameters. (l) Comparison of the extrapolated energy gap $\Delta_{0}$ in our Bi2267 with those measured in Bi2212\cite{SIdeta2010,IMVishik2012,MHashimoto2014} and Bi2223\cite{SIdeta2010,XYLuo2023}.
}
\end{figure*}
\clearpage


\vspace{3mm}

\noindent {\bf References}

\begin{thebibliography}{99}

\bibitem{PALee2006}
    P. A. Lee, N. Nagaosa and X. G. Wen.
    \newblock {Doping a Mott insulator: physics of high-temperature superconductivity}.
	\newblock {\em Review of Modern Physics} {\bf78}, 17 (2006).

\bibitem{BKeimer2015}
	B. Keimer, S. A. Kivelson, M. R. Norman, S. Uchida and J. Zaanen.
	\newblock {From quantum matter to high-temperature superconductivity in copper oxides}.
	\newblock {\em Nature} {\bf518}, 179 (2015).

\bibitem{DVaknin1987}
	D. Vaknin, S. K. Sinha, D. E. Moncton, D. C. Johnston, J. M. Newsam, C. R. Safinya and H. E. King, Jr.
	\newblock {Antiferromagnetism in $La_{2}CuO_{4-y}$}.
	\newblock {\em Physical Review Letters} {\bf58}, 2802 (1987).

\bibitem{JMTranquada1988}
	J. M. Tranquada, A. H. Moudden, A. I. Goldman, P. Zolliker, D. E. Cox, and G. Shirane, S. K. Sinha, D. Vaknin, D. C. Johnston, M. S. Alvarez, A. J. Jacobson, J. T. Lewandowski, and J. M. Newsam.
	\newblock {Antiferromagnetism in $YBa_{2}Cu_{3}O_{6+x}$
}.
	\newblock {\em Physical Review B} {\bf38}, 2477 (1988).

\bibitem{EManousakis1991}
    E. Manousakis.
    \newblock {The spin-$\frac{1}{2}$ Heisenberg antiferromagnet on a square lattice and its application to the cuprous oxides}.
	\newblock {\em Review of Modern Physics} {\bf63}, 1 (1991).

\bibitem{HEskes1993}
	H. Eskes and J. H. Jefferson.
	\newblock {Superexchange in the cuprates}.
	\newblock {\em Physical Review B} {\bf48}, 9788 (1993).

\bibitem{CWChu2015}
	C. W. Chu, L. Z. Deng and B. Lv.
	\newblock {Hole-doped cuprate high temperature superconductors}.
	\newblock {\em Physica C} {\bf514}, 290 (2015).

\bibitem{PCDai1999}
	P. C. Dai, H. A. Mook, S. M. Hayden, G. Aeppli, T. G. Perring, R. D. Hunt and F. Do$\breve{g}$an.
	\newblock {The magnetic excitation spectrum and thermodynamics of high-T$_{C}$ superconductors}.
	\newblock {\em Science} {\bf284}, 1344 (1999).

\bibitem{RColdea2001}
	R. Coldea, S. M. Hayden, G. Aeppli, T. G. Perring, C. D. Frost, T. E. Mason, S.-W. Cheong and Z. Fisk.
	\newblock {Spin waves and electronic interactions in $La_{2}CuO_{4}$}.
	\newblock {\em Physical Review Letters} {\bf86}, 5377 (2001).

\bibitem{HAMook2002}
	H. A. Mook, P. C. Dai and F. Do$\breve{g}$an.
	\newblock {Charge and Spin Structure in $YBa_{2}Cu_{3}O_{6.35}$}.
	\newblock {\em Physical Review Letters} {\bf88}, 097004 (2002).

\bibitem{SMHayden2004}
	S. M. Hayden, H. A. Mook, P. C. Dai, T. G. Perring and F. Do$\breve{g}$an.
	\newblock {The structure of the high-energy spin excitations in a high-transition-temperature superconductor}.
	\newblock {\em Nature} {\bf429}, 531 (2004).

\bibitem{SWakimoto2004}
	S. Wakimoto, H. Zhang, K. Yamada, I. Swainson, H. Kim and R. J. Birgeneau.
	\newblock {Direct relation between the low-energy spin excitations and superconductivity of overdoped high-T$_{C}$ superconductors}.
	\newblock {\em Physical Review Letters} {\bf92}, 217004 (2004).

\bibitem{BVignolle2007}
	B. Vignolle, S. M. Hayden, D. F. McMorrow, H. M. Ronnow, B. Lake, C. D. Frost and T. G. Perring.
	\newblock {Two energy scales in the spin excitations of the high-temperature superconductor $La_{2-x}Sr_{x}CuO_{4}$}.
	\newblock {\em Nature Physics} {\bf3}, 163 (2007).

\bibitem{JMTranquada2007}
	J. M. Tranquada.
	\newblock {Neutron scattering studies of antiferromagnetic correlations in cuprates}.
	\newblock in {\em Handbook of High-Temperature Superconductivity: Theory and Experiment} edited by J. R. Schrieffer (Springer, New York, 2007).

\bibitem{DJScalapino1986}
	D. J. Scalapino, E. Loh, Jr. and J. E. Hirsch.
	\newblock {d-wave pairing near a spin-density-wave instability}.
    \newblock {\em Physical Review B} {\bf34}, 8190 (1986).

\bibitem{PMonthoux1991}
	P. Monthoux, A. V. Balatsky and D. Pines.
	\newblock {Toward a theory of high-temperature superconductivity in the antiferromagnetically correlated cuprate oxides}.
	\newblock {\em Physical Review Letters} {\bf67}, 3448 (1991).

\bibitem{DJScalapino1995}
	D. J. Scalapino.
	\newblock {The case for $d_{x^{2}-y^{2}}$ pairing in the cuprate superconductors}.
	\newblock {\em Physics Reports} {\bf250}, 329 (1995).

\bibitem{TMoriya2000}
	T. Moriya and K. Ueda.
	\newblock {Spin fluctuations and high temperature superconductivity}.
	\newblock {\em Advances in Physics} {\bf49}, 555 (2000).

\bibitem{ArAbanov2008}
	Ar. Abanov, A. V. Chubukov and M. R. Norman.
	\newblock {Gap anisotropy and universal pairing scale in a spin-fluctuation model of cuprate superconductors}.
	\newblock {\em Physical Review B} {\bf78}, 220507 (2008).

\bibitem{DJScalapino2012}
	D. J. Scalapino.
	\newblock {A common thread: The pairing interaction for unconventional superconductors}.
	\newblock {\em Reviews of Modern Physics} {\bf84}, 1383 (2012).

\bibitem{ADamascelli2003}
	A. Damascelli, Z. Hussain and Z. X. Shen.
	\newblock {Angle-resolved photoemission studies of the cuprate superconductors}.
	\newblock {\em Reviews of Modern Physics} {\bf75}, 473 (2003).

\bibitem{TYoshida2003}
	T. Yoshida, X. J. Zhou, T. Sasagawa, W. L. Yang, P. V. Bogdanov, A. Lanzara, Z. Hussain, T. Mizokawa, A. Fujimori, H. Eisaki, Z.-X. Shen, T. Kakeshita and S. Uchida.
	\newblock {Metallic behavior of lightly doped $La_{2-x}Sr_{x}CuO_{4}$ with a Fermi surface forming an arc}.
	\newblock {\em Physical Review Letters} {\bf91}, 027001 (2003).

\bibitem{KMShen2004}
	  K. M. Shen, F. Ronning, D. H. Lu, W. S. Lee, N. J. C. Ingle, W. Meevasana, F. Baumberger, A. Damascelli, N. P. Armitage, L. L. Miller, Y. Kohsaka, M. Azuma, M. Takano, H. Takagi and Z. X. Shen.
	\newblock {Missing quasiparticles and the chemical potential puzzle in the doping evolution of the cuprate superconductors}.
	\newblock {\em Physical Review Letters} {\bf93}, 267002 (2004).

\bibitem{KMShen2005}
	  K. M. Shen, F. Ronning, D. H. Lu, F. Baumberger, N. J. C. Ingle, W. S. Lee, W. Meevasana, Y. Kohsaka, M. Azuma, M. Takano, H. Takagi and Z.-X. Shen.
	\newblock {Nodal quasiparticles and antinodal charge ordering in $Ca_{2-x}Na_{x}CuO_{2}Cl_{2}$}.
	\newblock {\em Science} {\bf307}, 901 (2005).

\bibitem{TYoshida2006}
	T. Yoshida, X. J. Zhou, K. Tanaka, W. L. Yang, Z. Hussain, Z.-X. Shen, A. Fujimori, S. Sahrakorpi, M. Lindroos, R. S. Markiewicz, A. Bansil, S. Komiya, Y. Ando, H. Eisaki, T. Kakeshita and S. Uchida.
	\newblock {Systematic doping evolution of the underlying Fermi surface of $La_{2-x}Sr_{x}CuO_{4}$}.
	\newblock {\em Physical Review B} {\bf74}, 224510 (2006).

\bibitem{MHashimoto2008}
	M. Hashimoto, T. Yoshida, H. Yagi, M. Takizawa, A. Fujimori, M. Kubota, K. Ono, K. Tanaka, D. H. Lu, Z.-X. Shen, S. Ono and Y. Ando.
    \newblock {Doping evolution of the electronic structure in the single-layer cuprate $Bi_{2}Sr_{2-x}La_{x}CuO_{6+\delta}$: Comparison with other single-layer cuprates}.
	\newblock {\em Physical Review B} {\bf77}, 094516 (2008).

\bibitem{YYPeng2013}
	Y. Y. Peng, J. Q. Meng, D. X. Mou, J. F. He, L. Zhao, Y. Wu, G. D. Liu, X. L. Dong, S. L. He, J. Zhang, X. Y. Wang, Q. J. Peng, Z. M. Wang, S. J. Zhang, F. Yang, C. T. Chen, Z. Y. Xu, T. K. Lee and X. J. Zhou.
	\newblock {Disappearance of nodal gap across the insulator-superconductor transition in a copper-oxide superconductor}.
	\newblock {\em Nature Communications} {\bf4}, 2459 (2013).

\bibitem{QGao2020}
	Q. Gao, L. Zhao, C. Hu, H. T. Yan, H. Chen, Y. Q. Cai, C. Li, P. Ai, J. Liu, J. W. Huang, H. T. Rong, C. Y. Song, C. H. Yin, Q. Y. Wang, Y. Huang, G. D. Liu, Z. Y. Xu and X. J. Zhou.
	\newblock {Electronic evolution from the parent Mott insulator to a superconductor in lightly hole-doped $Bi_{2}Sr_{2}CaCu_{2}O_{8+\delta}$}.
	\newblock {\em Chinese Physics Letters} {\bf37}, 087402 (2020).

\bibitem{JCCampuzano1999}
	  J. C. Campuzano, H. Ding, M. R. Norman, H. M. Fretwell, M. Randeria, A. Kaminski, J. Mesot, T. Takeuchi, T. Sato, T. Yokoya, T. Takahashi, T. Mochiku, K. Kadowaki, P. Guptasarma, D. G. Hinks, Z. Konstantinovic, Z. Z. Li and H. Raffy.
	\newblock {Electronic spectra and their relation to the $(\pi,\pi)$ collective mode in high-T$_{C}$ superconductors}.
	\newblock {\em Physical Review Letters} {\bf83}, 3709 (1999).

\bibitem{KTanaka2006}
    K. Tanaka, W. S. Lee, D. H. Lu, A. Fujimori, T. Fujii, Risdiana, I. Terasaki, D. J. Scalapino, T. P. Devereaux, Z. Hussain and Z.-X. Shen.
    \newblock {Distinct Fermi-momentum-dependent energy gaps in deeply underdoped Bi2212}.
	\newblock {\em Science} {\bf314}, 1910 (2006).

\bibitem{DLFeng2000}
	D. L. Feng, D. H. Lu, K. M. Shen, C. Kim, H. Eisaki, A. Damascelli, R. Yoshizaki, J.-i. Shimoyama, K. Kishio, G. D. Gu, S. Oh, A. Andrus, J. O$'$Donnell, J. N. Eckstein and Z. X. Shen.
	\newblock {Signature of superfluid density in the single-particle excitation spectrum of $Bi_{2}Sr_{2}CaCu_{2}O_{8+\delta}$}.
	\newblock {\em Science} {\bf289}, 434 (2000).

\bibitem{HDing2001}
	H. Ding, J. R. Engelbrecht, Z. Wang, J. C. Campuzano, S. C. Wang, H. B. Yang, R. Rogan, T. Takahashi, K. Kadowaki and D. G. Hinks.
	\newblock {Coherent quasiparticle weight and its connection to high-T$_{C}$ superconductivity from angle-resolved photoemission}.
	\newblock {\em Physical Review Letters} {\bf87}, 227001 (2001).

\bibitem{XJZhou2004}
X. J. Zhou, T. Yoshida, D.H. Lee, W. L. Yang, V. Brouet, F. Zhou, W. X. Ti, J. W. Xiong, Z. X. Zhao, T. Sasagawa, T. Kakeshita, H. Eisaki, S. Uchida, A. Fujimori, Z. Hussain and Z. X. Shen.
\newblock {Dichotomy between nodal and antinodal quasiparticles in underdoped $(La_{2-x}Sr_{x})CuO_{4}$ superconductors}.
\newblock {\em Physical Review Letters} {\bf92}, 187001 (2004).

\bibitem{TDahm2009}
T. Dahm, V. Hinkov, S. V. Borisenko, A. A. Kordyuk, V. B. Zabolotnyy, J. Fink, B. Buechner, D. J. Scalapino, W. Hanke and B. Keimer.
\newblock {Strength of the spin-fluctuation-mediated pairing interaction in a high-temperature superconductor}.
\newblock {\em Nature Physics} {\bf5}, 217 (2009).

\bibitem{ZWang2023}
    Z. C. Wang, C. W. Zou, C. T. Lin, X. Y. Luo, H. T. Yan, C. H. Yin, Y. Xu, X. J. Zhou, Y. Y. Wang and J. Zhu.
    \newblock {Correlating the charge-transfer gap to the maximum transition temperature in $Bi_{2}Sr_{2}Ca_{n-1}Cu_{n}O_{2n+4+\delta}$}.
	\newblock {\em Science} {\bf381}, 227 (2023).

\bibitem{SKunisada2020}
    S. Kunisada, S. Isono, Y. Kohama, S. Sakai, C. Bareille, S. Sakuragi, R. Noguchi, K. Kurokawa, K. Kuroda, Y. Ishida, S. Adachi, R. Sekine, T. K. Kim, C. Cacho, S Shin, T. Tohyama, K. Tokiwa and T. Kondo.
    \newblock {Observation of small Fermi pockets protected by clean $CuO_{2}$ sheets of a high-$T_{C}$ superconductor}.
	\newblock {\em Science} {\bf369}, 833 (2020).

\bibitem{KKurokawa2023}
     K. Kurokawa, S. Isono, Y. Kohama, S. Kunisada, S. Sakai, R. Sekine, M. Okubo, M. D. Watson, T. K. Kim, C. Cacho, S. Shin, T. Tohyama, K. Tokiwa and T. Kondo.
    \newblock {Unveiling phase diagram of the lightly doped high-$T_{C}$ cuprate superconductors with disorder removed}.
	\newblock {\em Nature Communications} {\bf14}, 4064 (2023).

\bibitem{TKondo2007}
	  T. Kondo, T. Takeuchi, A. Kaminski, S. Tsuda and S. Shin.
	\newblock {Evidence for two energy scales in the superconducting state of optimally doped $(Bi,Pb)_{2}(Sr,La)_{2}CuO_{6+\delta}$}.
	\newblock {\em Physical Review Letters} {\bf98}, 267004 (2007).

\bibitem{TKondo2009}
	 T. Kondo, R. Khasanov, T. Takeuchi, J. Schmalian and A. Kaminski.
	\newblock {Competition between the pseudogap and superconductivity in the high-$T_{C}$ copper oxides}.
	\newblock {\em Nature} {\bf457}, 296 (2009).

\bibitem{YXZhang2016}
	 Y. X. Zhang, C. Hu, Y. Hu, L. Zhao, Y. Ding, X. Sun, A. J. Liang, Y. Zhang, S. L. He, D. F. Liu, L. Yu, G. D. Liu, X. L. Dong, G. D. Gu, C. T. Chen, Z. Y. Xu and X. J. Zhou.
	\newblock {In situ carrier tuning in high temperature superconductor $Bi_{2}Sr_{2}CaCu_{2}O_{8+\delta}$ by potassium deposition}.
	\newblock {\em Science Bulletin} {\bf61}, 1037 (2016).

\bibitem{IKDrozdov2018}
	I. K. Drozdov, I. Pletikosi$\acute{c}$, C.-K. Kim, K. Fujita, G. D. Gu, J. C. S$\acute{e}$amus Davis, P. D. Johnson, I. Bo$\breve{z}$ovi$\acute{c}$ and T. Valla.
	\newblock {Phase diagram of $Bi_{2}Sr_{2}CaCu_{2}O_{8+\delta}$ revisited}.
	\newblock {\em Nature Communications} {\bf9}, 5210 (2018).

\bibitem{YHe2018}
	Y. He, M. Hashimoto, D. Song, S.-D. Chen, J. He, I. M. Vishik, B. Moritz, D.-H. Lee, N. Nagaosa, J. Zaanen, T. P. Devereaux, Y. Yoshida, H. Eisaki, D. H. Lu and Z.-X. Shen.
	\newblock {Rapid change of superconductivity and electron-phonon coupling through critical doping in Bi-2212}.
	\newblock {\em Science} {\bf362}, 62 (2018).

\bibitem{YGZhong2018}
	Y. G. Zhong, Y. M. Chen, J. Y. Guan, J. Zhao, Z. C. Rao, C. Y. Tang, H. J. Liu, Y. J. Sun and H. Ding.
	\newblock {Extraction of tight binding parameters from in-situ ARPES on the continuously doped surface of cuprates}.
	\newblock {\em Science China-Physics, Mechanics and Astronomy.} {\bf61}, 127403 (2018).

\bibitem{YDing2019}
	Y. Ding, L. Zhao, H. T. Yan, Q. Gao, J. Liu, C. Hu, J. W. Huang, C. Li, Y. Xu, Y. Q. Cai, H. T. Rong, D. S. Wu, C. Y. Song, H. X. Zhou, X. L. Dong, G. D. Liu, Q. Y. Wang, S. J. Zhang, Z. M. Wang, F. F. Zhang, F. Yang, Q. J. Peng, Z. Y. Xu, C. T. Chen and X. J. Zhou.
	\newblock {Disappearance of superconductivity and a concomitant Lifshitz transition in heavily overdoped $Bi_{2}Sr_{2}CuO_{6}$ superconductor revealed by angle-resolved photoemission spectroscopy}.
	\newblock {\em Chinese Physics Letters} {\bf36}, 017402 (2019).

\bibitem{PAi2019}
	P. Ai, Q. Gao, J. Liu, Y. X. Zhang, C. Li, J. W. Huang, C. Y. Song, H. T. Yan, L. Zhao, G. D. Liu, G. D. Gu, F. F. Zhang, F. Yang, Q. J. Peng, Z. Y. Xu, and X. J. Zhou.
	\newblock {Distinct superconducting gap on two bilayer-split Fermi surface sheets in $Bi_{2}Sr_{2}CaCu_{2}O_{8+\delta}$ superconductor}.
	\newblock {\em Chinese Physics Letters} {\bf36}, 067402 (2019).

\bibitem{TValla2020}
	T. Valla, I. K. Drozdov and G.D. Gu.
	\newblock {Disappearance of superconductivity due
 to vanishing coupling in the overdoped $Bi_{2}Sr_{2}CaCu_{2}O_{8+\delta}$}.
	\newblock {\em Nature Communications} {\bf11}, 569 (2020).

\bibitem{TValla2021}
	T. Valla, P. Pervan, I. Pletikosi$\acute{c}$, I. K. Drozdov, A. K. Kundu, Z. Wu and G. D. Gu.
	\newblock {Hole-like Fermi surface in the overdoped non-superconducting $Bi_{1.8}Pb_{0.4}Sr_{2}CuO_{6+\delta}$}.
	\newblock {\em Europhysics Letters} {\bf134}, 17002 (2021).

\bibitem{RSMarkiewicz2005}
	R. S. Markiewicz, S. Sahrakorpi, M. Lindroos, H. Lin and A. Bansil.
	\newblock {One-band tight-binding model parametrization of the high-$T_{c}$ cuprates including the effect of $k_{z}$ dispersion}.
	\newblock {\em Physical Review B} {\bf72}, 054519 (2005).

\bibitem{HMukuda2012}
	H. Mukuda, S. Shimizu, A. Iyo, and Y. Kitaoka.
	\newblock {High-$T_{C}$ superconductivity and antiferromagnetism in multilayered copper oxides -a new paradigm of superconducting mechanism}.
	\newblock {\em Journal of the Physical Society of Japan} {\bf81}, 011008 (2012).

\bibitem{MMori2006}
	M. Mori, T. Tohyama and S. Maekawa.
	\newblock {Charge imbalance effects on interlayer hopping and Fermi surfaces in multilayered high-T$_{C}$ cuprates}.
	\newblock {\em Journal of the Physical Society of Japan} {\bf75}, 034708 (2006).

\bibitem{HChen2026}
    H. Chen, J. M. Shi, Y. H. Li, X. Y. Luo, Y. W. Chen, C. H. Yin, Y. J. Shu, J. X. Zhang, T. M. Miao, B. Liang, W. P. Zhu, N. Cai, X. L. Ren, C. T. Lin, S. J. Zhang, Z. M. Wang, F. F. Zhang, F. Yang, Q. J. Peng, Z. Y. Xu, G. D. Liu, H. Q. Mao, X. T. Li, T. Xiang, L. Zhao and X. J. Zhou.
	\newblock {Persistent Fermi pockets and robust electron pairing in lightly doped CuO$_2$ planes of cuprate superconductors}. arXiv:2604.23162 (2026).

\bibitem{SKunisada2017}
	S. Kunisada, S. Adachi, S. Sakai, N. Sasaki, M. Nakayama, S. Akebi, K. Kuroda, T. Sasagawa, T. Watanabe, S. Shin, and T. Kondo.
	\newblock {Observation of Bogoliubov band hybridization in the optimally doped trilayer $Bi_{2}Sr_{2}Ca_{2}Cu_{3}O_{10+\delta}$}.
	\newblock {\em Physical Review Letters} {\bf119}, 217001 (2017).

\bibitem{SIdeta2021}
	S. Ideta, S. Johnston, T. Yoshida, K. Tanaka, M. Mori, H. Anzai, A. Ino, M. Arita, H. Namatame, M. Taniguchi, S. Ishida, K. Takashima, K. M. Kojima, T. P. Devereaux, S. Uchida, and A. Fujimori.
	\newblock {Hybridization of Bogoliubov quasiparticles between adjacent CuO$_{2}$ layers in the triple-layer cuprate $Bi_{2}Sr_{2}Ca_{2}Cu_{3}O_{10+\delta}$ studied by angle-resolved photoemission spectroscopy}.
	\newblock {\em Physical Review Letters} {\bf127}, 217004 (2021).

\bibitem{XYLuo2023}
	 X. Y. Luo, H. Chen, Y. H. Li, Q. Gao, C. H. Yin, H. T. Yan, T. M. Miao, H. L. Luo, Y. J. Shu, Y. W. Chen, C. T. Lin, S. J. Zhang, Z. M. Wang, F. F. Zhang, F. Yang, Q. J. Peng, G. D. Liu, L. Zhao, Z. Y. Xu, T. Xiang and X. J. Zhou.
	\newblock {Electronic origin of high superconducting critical temperature in trilayer cuprates}.
	\newblock {\em Nature Physics} {\bf19}, 1841 (2023).

\bibitem{MHashimoto2014}
	 M. Hashimoto, I. M. Vishik, R. H. He, T. P. Devereaux and Z. X. Shen.
	\newblock {Energy gaps in high-transition-temperature cuprate superconductors}.
	\newblock {\em Nature Physics} {\bf10}, 483 (2014).

\bibitem{SIdeta2010}
	S. Ideta, K. Takashima, M. Hashimoto, T. Yoshida, A. Fujimori, H. Anzai, T. Fujita, Y. Nakashima, A. Ino, M. Arita, H. Namatame, M. Taniguchi, K. Ono, M. Kubota, D. H. Lu, Z.-X. Shen, K. M. Kojima, and S. Uchida.
	\newblock {Enhanced superconducting gaps in the trilayer high-temperature $Bi_{2}Sr_{2}Ca_{2}Cu_{3}O_{10+\delta}$ cuprate superconductor}.
	\newblock {\em  Physical Review Letters} {\bf104}, 227001 (2010).

\bibitem{YXu2021}
	 Y. Xu, H. T. Rong, Q. Y. Wang, D. S. Wu, Y. Hu, Y. Q. Cai, Q. Gao, H. T. Yan, C. Li, C. H. Yin, H. Chen, J. W. Huang, Z. H. Zhu, Y. Huang, G. D. Liu, Z. Y. Xu, L. Zhao and X. J. Zhou.
	\newblock {Spectroscopic evidence of superconductivity pairing at 83 K in single-layer FeSe/SrTiO$_{3}$ films}.
	\newblock {\em Nature Communications} {\bf12}, 2840 (2021).

\bibitem{SDChen2022}
	 S. D. Chen, M. Hashimoto, Y. He, D. Song, J. F. He, Y. F. Li, S. Ishida, H. Eisaki, J. Zaanen, T. P. Devereaux, D. H. Lee, D. H. Lu and Z. X. Shen.
	\newblock {Unconventional spectral signature of T$_{C}$ in a pure d-wave superconductor}.
	\newblock {\em Nature} {\bf601}, 562 (2022).

\bibitem{HNarita1992}
H. Narita, T. Hatano and K. Nakamura.
\newblock {Synthesis and characterization of Bi$_2$Sr$_2$Ca$_{n-1}$Cu$_n$O$_y$ ($n$ = 1-7) thin films grown by off-axis, three target magnetron sputtering}.
\newblock {\em Journal of Applied Physics} {\bf72}, 5778 (1992).

\bibitem{HMatsui2005}
H. Matsui, K. Terashima, T. Sato, T. Takahashi, S.-C. Wang, H.-B. Yang, H. Ding, T. Uefuji and K. Yamada.
\newblock {Angle-Resolved Photoemission Spectroscopy of the Antiferromagnetic Superconductor $Nd_{1.87}Ce_{0.13}CuO_{4}$: Anisotropic spin-correlation gap, pseudogap, and the induced quasiparticle mass enhancement}.
\newblock {\em Physical Review Letters} {\bf94}, 047005 (2005).

\bibitem{JHe2019}
J. He, C. R. Rotundu, M. S. Scheurer, Y. He, M. Hashimoto, K.-J. Xu, Y. Wang, E. W. Huang, T. Jia, S. Chen, B. Moritz, D. Lu, Y. S. Lee, T. P. Devereaux and Z.-X. Shen,
\newblock {Fermi surface reconstruction in electron-doped cuprates without antiferromagnetic long-range order}.
\newblock {\em Proceedings of the National Academy of Sciences} {\bf116}, 3449 (2019).

\bibitem{JRSchrieffer1989}
	 J. R. Schrieffer, X. G. Wen and S. C. Zhang.
	\newblock {Dynamic spin fluctuations and the bag mechanism of high-T$_{C}$ superconductivity}.
	\newblock {\em Physical Review B} {\bf39}, 11663 (1989).

\bibitem{AVChubukov1992}
	 A. V. Chubukov and D. M. Frenkel.
	\newblock {Renormalized perturbation theory of magnetic instabilities in the two-dimensional Hubbard model at small doping}.
	\newblock {\em Physical Review B} {\bf46}, 11884 (1992).

\bibitem{NBulut1994}
	 N. Bulut, D. J. Scalapino and S. R. White.
	\newblock {Electronic properties of the insulating half-filled hubbard model}.
	\newblock {\em Physical Review Letters} {\bf73}, 748 (1994).

\bibitem{TXiang1996}
	 T. Xiang and J. M. Wheatley.
	\newblock {Quasiparticle energy dispersion in doped two-dimensional quantum antiferromagnets}.
	\newblock {\em Physical Review B} {\bf54}, R12653 (1996).

\bibitem{CKusko2002}
	 C. Kusko, R. S. Markiewicz, M. Lindroos and A. Bansil.
	\newblock {Fermi surface evolution and collapse of the Mott pseudogap in $Nd_{2-x}Ce_{x}CuO_{4\pm\delta}$}.
	\newblock {\em Physical Review B} {\bf66}, 140513 (2002).

\bibitem{VJEmery1995}
V. J. Emery and S. A. Kivelson.
\newblock {Importance of phase fluctuations in superconductors with small superfluid density}.
\newblock {\em Nature} {\bf374}, 434 (1995).

\bibitem{UChatterjee2010}
	U. Chatterjee, M. Shi, D. Ai, J. Zhao, A. Kanigel, S. Rosenkranz, H. Raffy, Z. Z. Li, K. Kadowaki, D. G. Hinks, Z. J. Xu, J. S.Wen, G. Gu, C. T. Lin, H. Claus, M. R. Norman, M. Randeria and J. C. Campuzano.
	\newblock {Observation of a d-wave nodal liquid in highly underdoped $Bi_{2}Sr_{2}CaCu_{2}O_{8+\delta}$}.
	\newblock {\em  Nature Physics} {\bf6}, 99 (2010).

\bibitem{IMVishik2012}
    I. M. Vishik, M. Hashimoto, R. H. He, W. S. Lee,b, F. Schmitt, D. H. Lu, R. G. Moore, C. Zhang, W. Meevasana, T. Sasagawa, S. Uchida, K. Fujita, S. Ishida, M. Ishikado, Y. Yoshida, H. Eisaki, Z. Hussain, T. P. Devereaux and Z. X. Shen.
\newblock {Phase competition in trisected superconducting dome}.
\newblock {\em Proceedings of the National Academy of Sciences of the United States of America} {\bf109}, 18332 (2012).

\bibitem{BLiang2002}
	B. Liang, C. T. Lin, P. Shang and G. Yang.
	\newblock {Single crystals of triple-layered cuprates $Bi_{2}Sr_{2}Ca_{2}Cu_{3}O_{10+\delta}$: growth, annealing and characterization}.
	\newblock {\em Physica C} {\bf383}, 75 (2002).

\bibitem{CTLin2002}
	C. T. Lin and B. Liang.
	\newblock {Growth of a hard-grown single crystal-$Bi_{2}Sr_{2}Ca_{2}Cu_{3}O_{10+\delta}$}.
	\newblock in {\em New Trends in Superconductivity} edited by J.F. Annett and S. Kruchinin (Kluwer
Academic Publishers, The Netherlands, 2002).

\bibitem{XJZhou2008GDLiu}
	G. D. Liu, G. L. Wang, Y. Zhu, H. B. Zhang, G. C. Zhang, X. Y. Wang, Y. Zhou, W. T. Zhang, H. Y. Liu, L. Zhao, J. Q. Meng, X. L. Dong, C. T. Chen, Z. Y. Xu, and X. J. Zhou.
	\newblock {Development of a vacuum ultraviolet laser-based angle-resolved photoemission system with a superhigh energy resolution better than 1meV}.
	\newblock {\em Review of Scientific Instruments} {\bf79}, 023105 (2008).

\bibitem{WTZhang2018XJZhou}
	X. J. Zhou, S. L. He, G. D. Liu, L. Zhao, L. Yu, and W. T. Zhang.
	\newblock {New developments in laser-based photoemission spectroscopy and its scientific applications: a key issues review}.
	\newblock {\em Reports on Progress in Physics} {\bf81}, 062101 (2018).

\end{thebibliography}
\end{document}